\documentclass[journal,showpacs]{IEEEtran}
\usepackage{amsmath,amssymb,graphicx}
\usepackage{multirow}
\usepackage{epsf}
\usepackage{comment}

\usepackage{CJK}
\usepackage{amsmath}
\usepackage{indentfirst}%
\usepackage{graphicx}
\usepackage{float}
\usepackage{subcaption}
\usepackage{caption}

\begin{document}
%

\title{Secret key distillation over realistic satellite-to-satellite free-space   channel: \\exclusion zone analysis}
%
%

\author{Ziwen Pan and Ivan B. Djordjevic %
}

\markboth{Journal of \LaTeX\ Class Files,~Vol.~14, No.~18, August~2016}%
{Shell \MakeLowercase{\textit{et al.}}: Bare Demo of IEEEtran.cls for IEEE Journals}


\maketitle

\begin{abstract}



Quantum cryptography studies the unconditional information security against an all-powerful eavesdropper in secret key distillation. However the assumption of an omnipotent eavesdropper is too strict for some realistic implementations. In this paper we study the realistic application model of secret key distillation over satellite-to-satellite free space  channel 
in which we impose a reasonable restriction on the eavesdropper by setting an exclusion zone around the legitimate receiver as a defense strategy. 
We first study the case where the eavesdropper's aperture size is unlimited so her power is only restricted by the exclusion zone. Then we limit Eve's aperture to a finite size and study the straightforward case when her aperture is in the same plane of Bob, investigating how an exclusion zone can help improve security. Correspondingly, we determine the secret key rate lower bounds as well as upper bounds. Further more, we also apply our results on specific discrete variable (DV) and continuous variable (CV) protocols for comparison. 
We show that by putting reasonable restrictions on the eavesdropper through the realistic assumptions of an inaccessible exclusion zone, we can increase the key rate in comparison to those without and do so with relatively lower transmission frequency. We conclude that this model is suitable for extended analysis in many light gathering scenarios and for different carrier wavelengths.

\end{abstract}




\section{Introduction}

\IEEEPARstart{T}heoretically, quantum cryptography promises unconditional informational security. In 1984, Charles Bennett and Gilles Brassard developed a quantum key distribution (QKD) scheme, BB84~\cite{bennet1984proceedings},  with its security guaranteed by the no-cloning theorem and one-time pad encryption. This was the the first discrete variable QKD (DV-QKD) protocol, which uses single photons in transmission. This has put forward rigorous conditions for realistic implementations.

Nowadays as the need for security in communication has vastly grown, realistic implementations of QKD in applicable scenarios have been more and more important. Thus, thanks to its advantages in experimental implementation, continuous variable QKD (CV-QKD) has become an attractive field, e.g., protocols based on coherent laser light and heterodyne detection~\cite{LPFPP18, DL15}. However, most research around the security of QKD has assumed that the eavesdropper (Eve) has access to any operation that is allowed by physics law, which is not going to be very common in the near future. We have shown the security analysis of realistic secret key distillation with achievable rate calculation in~\cite{pan2019secret,8849223}. Certain restrictions to Eve's collecting ability were considered such as an exclusion zone around the legitimate receiver or a limited aperture size of Eve's receiver in a wireless channel.

In recent years, the capacity and security of communication between satellites have become relevant with the fast development of satellite-based free-space  communications. Since the work on satellite-to-ground quantum key distribution (QKD) in 2017~\cite{liao2017satellite}, interests have been rising surrounding free-space secret key distillation for satellites. 
In this paper we 
look into the typical scenario where eavesdropper's (Eve's) collecting ability is limited by an exclusion zone around the legitimate receiver. In Sec.~\ref{ExZo} we  introduce and analyze the so-called "exclusion zone" scenario  where Eve cannot eavesdrop without alerting the communicating parties in  
certain region near Bob. We start with presenting the input power dependency with respect to different reconciliation efficiencies and then optimize the input power to show secure key rate lower bounds and upper bounds as functions of both transmission distance and center frequency. We also incorporate these results with specific continuous-variable and discrete-variable protocols for comparison.
Then in Sec.~\ref{CombinedScenario} we look into the scenario where Eve's collecting ability is limited by the size of her aperture and study how an exclusion zone can increase security as a defense strategy by presenting lower bounds and upper bounds with respect to different radius of the exclusion zone.
We show that an exclusion zone can increase the secure key rate lower bounds and act as a valuable defense strategy, especially when the eavesdropper's aperture size is limited. 
We demonstrate that for certain well-chosen parameters we can achieve significantly higher SKR lower bounds compared to traditional unrestricted Eve scenario.

\section{Exclusion zone scenario}\label{ExZo}

\begin{figure}[H] 
\centering{\includegraphics[height=12pc]{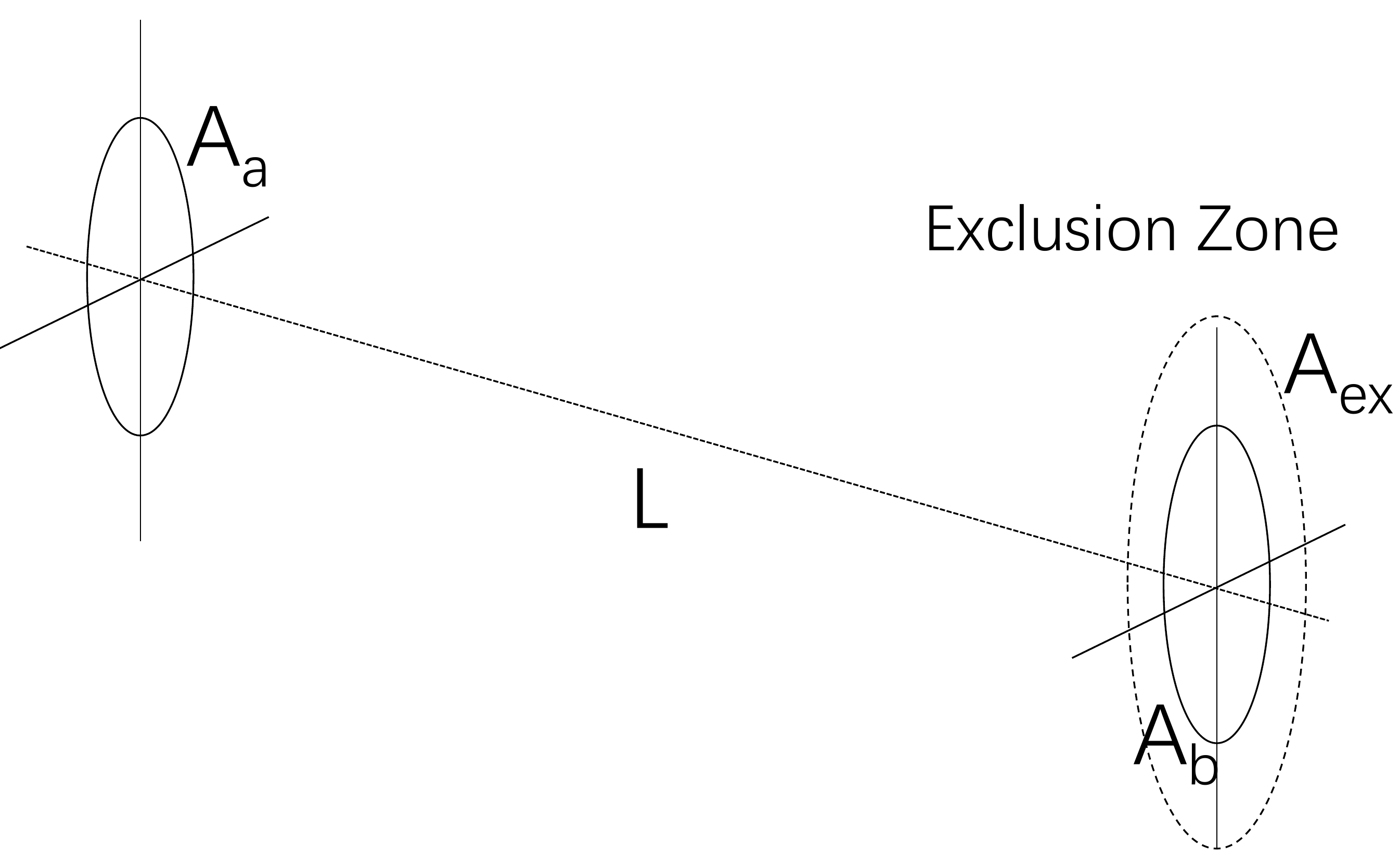}}
\caption{Geometric setup of exclusion zone scenario. $A_a$ is the transmitting aperture (Alice) area and $A_b$ is the receiving aperture (Bob) area. $L$ is the  transmission distance between Alice and Bob. The exclusion zone area  is denoted as $A_{ex}$ which is a circular area centered at Bob's aperture. \label{FSO}}
\end{figure}

\noindent In this section we will evaluate the exclusion zone scenario 
as is illustrated in Fig.~\ref{FSO}, 
which is called here the "exclusion zone" meaning that Eve cannot collect photons in this zone 
without being noticed by Alice and Bob. In satellite based secure communication, setting an "exclusion zone" is one of the most straightforward methods to improve security as this effectively decreases Eve's collecting ability. Here we assume that Eve can collect all photons 
outside of the exclusion zone.  

As illustrated in Fig.~\ref{FSO}, we assume the area of transmitter aperture is $A_a$ (Alice) and the area of receiver aperture is $A_b$ (Bob) with the area of so-called "exclusion zone" (dashed line) denoted as $A_{ex}$. Once more, we define the "exclusion zone" as the area/space in which Eve is not able to collect photons that end in this zone without alerting both communication parties. According to some similar analysis in \cite{shapiro2005ultimate,giovannetti2004classical}, if the frequency used $\omega$ is restriced to $0\leq\omega\ll\omega_0=2\pi cL/\sqrt{A_aA_b}$, then the Alice-to-Bob transmissivity $\eta$ at frequency $\omega$ is denoted as $\eta(\omega)=(\omega/\omega_0)^2\ll1$.

Applying this to our analysis, we will have:
\begin{align}
\omega_0&=2\pi cL/\sqrt{A_aA_b}\label{ExclusionZoneOmega0}\\
\eta(\omega)&=(\omega/\omega_0)^2\\
\omega_{0Ex}&=2\pi cL/\sqrt{A_aA_{ex}}\label{ExclusionZoneOmega0Ex}\\
\eta_{AEx}&=(\omega/\omega_{0Ex})^2\\
\eta_{AE}&=1-\eta_{AEx}=\left(1-\eta(\omega)\right)\kappa(\omega)\label{ExclusionZoneEtaEve}
\end{align}
Here $\kappa$ is the restriction factor~\cite{pan2019secret} on Eve denoting how much power she can collect from the part that isn't collected by Bob. Thus the transmissivity of Alice to Eve ($\eta_{AE}$) can be denoted as $\left(1-\eta(\omega)\right)\kappa(\omega)$ in Eq.~(\ref{ExclusionZoneEtaEve}). Eq.~(\ref{ExclusionZoneOmega0Ex}) gives us $\omega_{0Ex}$ which substitutes $A_b$ with $A_{ex}$ in Eq.~(\ref{ExclusionZoneOmega0}). This gives us Alice-to-exclusion zone transmissivity ($\eta_{AEx}$) assuming there is a virtual receiver covering the entire exclusion zone. Then assuming Eve has an infinite sized aperture, Alice-to-Eve transmissivity ($\eta_{AE}$) can be calculated by Eq.(\ref{ExclusionZoneEtaEve}).    

Also for noise frequency dependence we have the black body radiation equation:
\begin{equation}
n_e=\frac{1}{e^{\frac{hf}{kT}}-1}\label{blackradia},
\end{equation}
where $n_e$ is the mean photon number per mode for the thermal noise, $h=6.626*10^{-34}$m$^2$kg/s is the Planck constant, $k=1.38064852*10^{-23}$m$^2$kg/(Ks$^2$) is the Boltzmann constant, $T=3$K is the space temperature and $f$ is the center frequency in Hz that we use in transmission.

\begin{figure}[htbp]
\centering
\centering
\includegraphics[width=8.8cm]{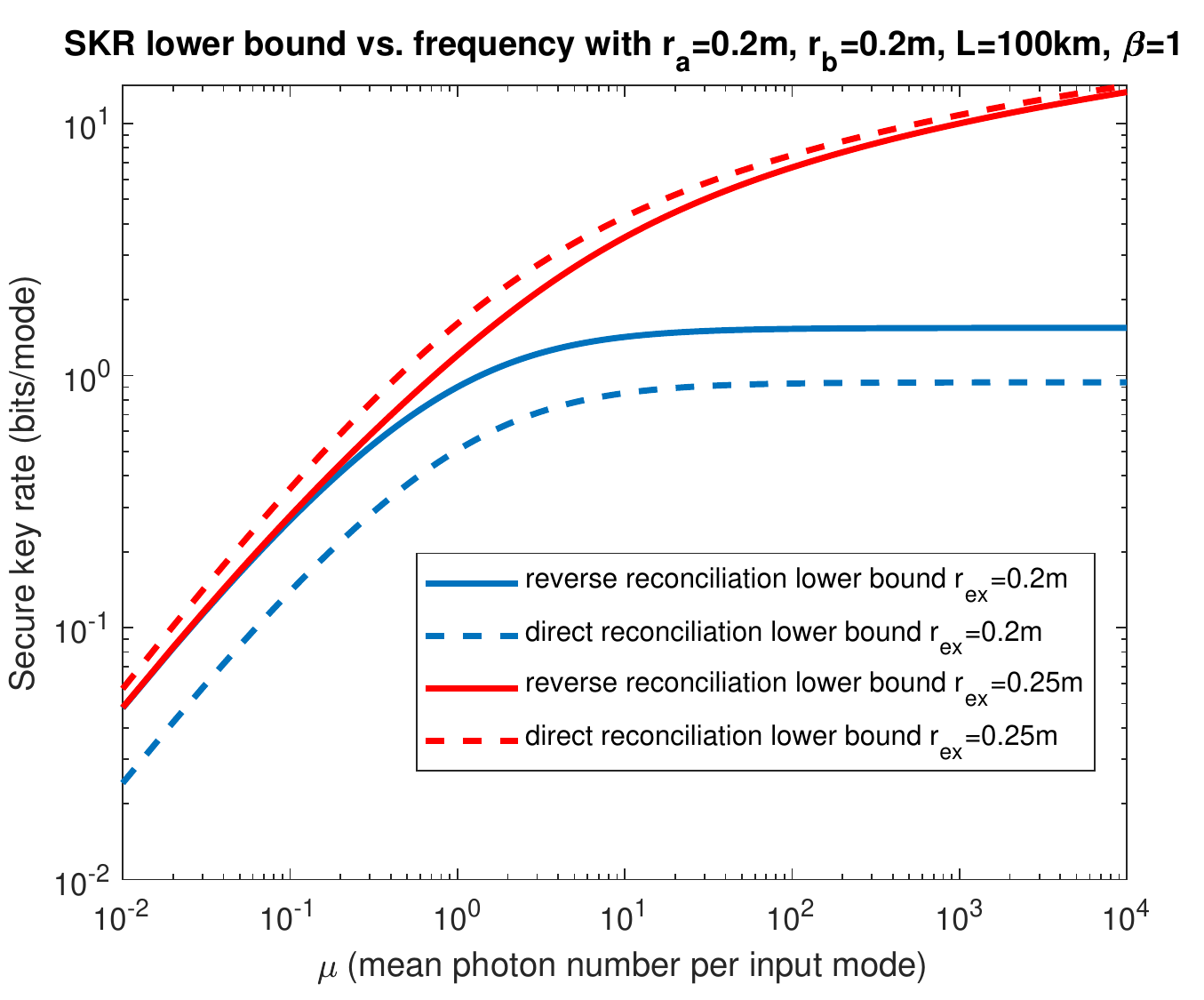}
\caption{SKR lower bound vs. input power. 
The transmission distance is 100km. Radius of exclusion zone ($r_{ex}$) is used as a parameter. Reconciliation efficiency $\beta$ is set to 1. Wavelength $\lambda$ is set to 1550nm. \label{ExZo_Index202005012254_cropped}}
\end{figure}

Recall from \cite{pan2019secret} the lower bound for direct ($K_\rightarrow$) and reverse ($K_\leftarrow$) reconciliation respectively in a quantum thermal noise wiretap channel:
\begin{align}
    K_\rightarrow &\geq \beta g\left(n_e(1-\eta)+\eta\mu\right)-\sum_i g\left(\frac{\nu^{ER}_i-1}{2}\right)\nonumber\\
&-\beta g\left(n_e(1-\eta)\right)+g\left(n_e(1-\eta\kappa)\right)\label{LBDmu}.\\
K_\leftarrow &\geq \beta g(\mu)-\sum_i g\left(\frac{\nu^{ER}_i-1}{2}\right)\nonumber\\
&-\beta g\left(\mu-\frac{\eta\mu(1+\mu)}{1+n_e-n_e\eta+\eta\mu}\right)+\sum_i g\left(\frac{\nu^{ER}_{y_i}-1}{2}\right)\label{LBRmu}.\\
g(x)&=(x+1)\log_2(x+1)-x\log_2(x).\label{thentropy}
\end{align}
where $\mu$ is the mean photon number per input mode from Alice to Bob, $\beta\in (0,1]$ is the reconciliation efficiency. In Fig.~\ref{ExZo_Index202005012254_cropped} we  plot direct (dashed curves) and reverse (solid curves) SKR lower bounds as functions of input power $\mu$ with different radius $r_{ex}$ of $A_{ex}$. Here we 
set transmission wavelength at 1550nm. We use blue curves to denote the case with no exclusion zone ($r_{ex}=r_b$) whereas red curves denotes the case with an exclusion zone ($r_{ex}>r_b$). The transmission distance is set as 100km and the reconciliation efficiency is set as $\beta=1$. 

Here in Fig.~\ref{ExZo_Index202005012254_cropped} we can see that although SKR lower bounds for both reconciliation schemes saturate as input power is sufficiently large when no exclusion zone is set ($r_{ex}=r_b$), we can exceed this saturation threshold by setting an exclusion zone (such as the red curves here with $r_{ex}=r_b+5~\text{cm}$) as this effectively diminishes Eve's collecting ability so that higher input power would benefit Bob more than Eve. Also it's worth noticing in Fig.~\ref{ExZo_Index202005012254_cropped} that an exclusion zone actually benefits direct reconciliation scheme more than reverse reconciliation. This is because setting an exclusion zone only diminishes Eve's collecting ability ($\kappa$) with no impact on Bob's receiving ability ($\eta$) and from~\cite{pan2019secret} we know that direct reconciliation exceeds reverse reconciliation when $\kappa$ is small with fixed $\eta$. 

In Fig.~\ref{ExZo_Index202005012254_cropped} we can also see that here SKR lower bounds are non-decreasing with increasing input power, meaning that when reconciliation efficiency is perfect ($\beta=1$), the optimal input power is infinity. Below in Fig.~\ref{ExZo_Index202005012254_beta95_cropped} we show that optimal input power is finite when reconciliation efficiency is imperfect ($\beta=0.95$). By setting an exclusion zone ($r_{ex}=r_b+5\text{cm}$ in Fig.~\ref{ExZo_Index202005012254_beta95_cropped}) we are able to achieve higher SKR exceeding this optimal input power.

\begin{figure}[htbp]
\centering
\centering
\includegraphics[width=8.8cm]{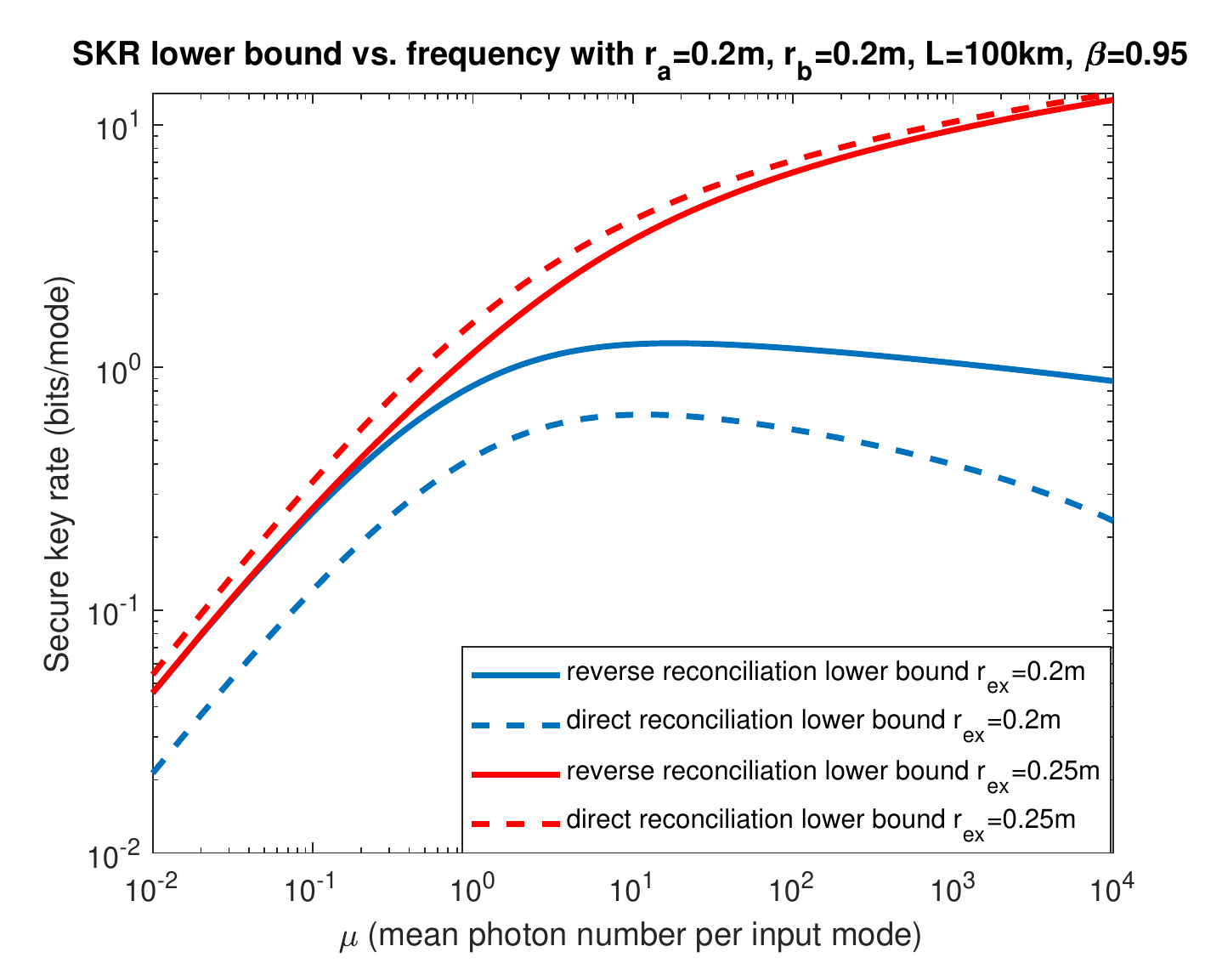}
\caption{SKR lower bound vs. input power. 
The transmission distance is 100km. Radius of exclusion zone ($r_{ex}$) is used as a parameter. Reconciliation efficiency $\beta$ is set to 0.95. Wavelength $\lambda$ is set to 1550nm. \label{ExZo_Index202005012254_beta95_cropped}}
\end{figure}
\begin{figure}[htbp]
\centering
\centering
\includegraphics[width=8.8cm]{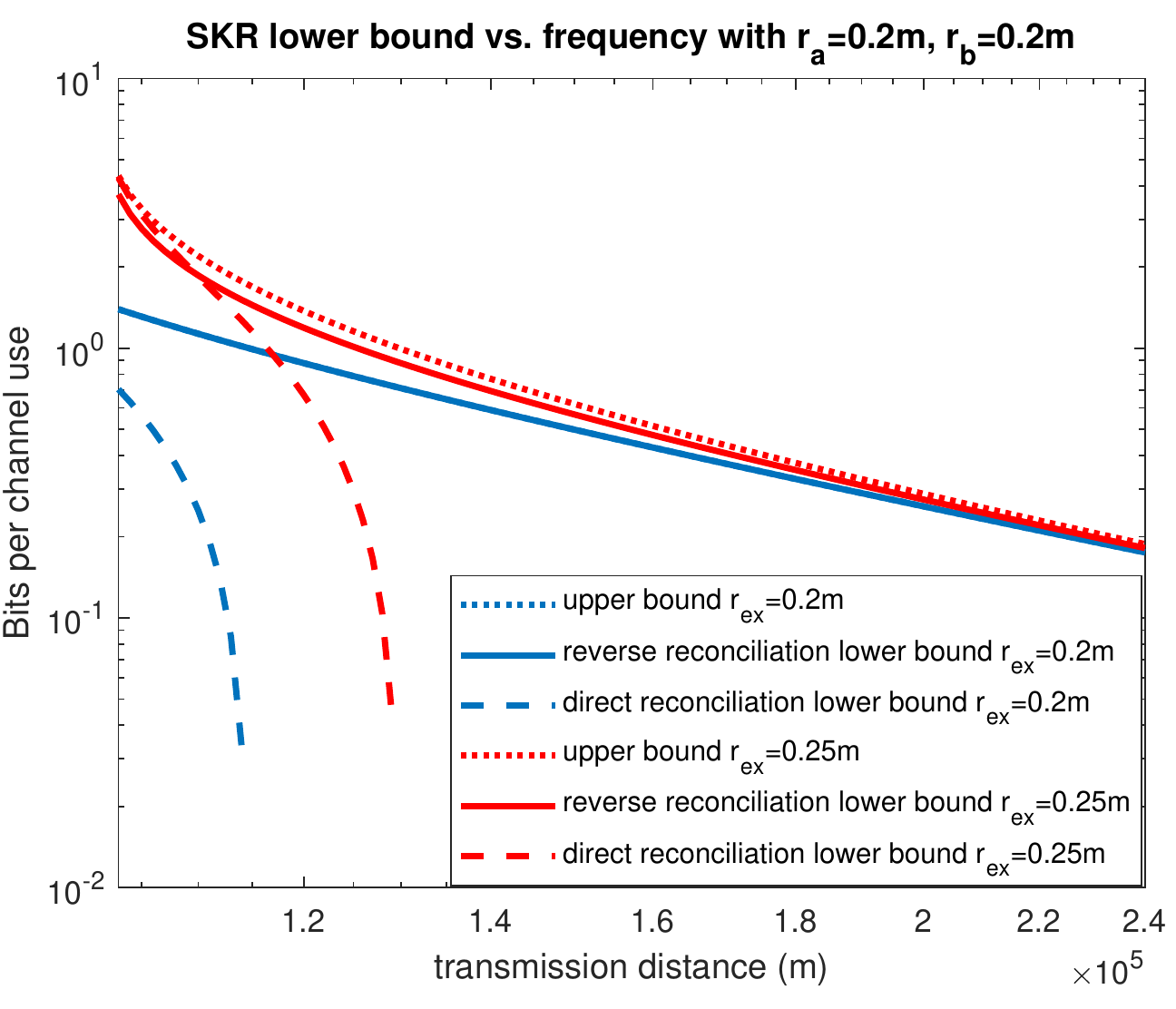}
\caption{SKR lower bound vs. transmission distance. 
Radius of exclusion zone ($r_{ex}$) 
is used as a parameter. Wavelength $\lambda$ is set to 1550nm. Reconciliation efficiency is perfect ($\beta=1$) and the input power is set to infinity.\label{ExclusionZonePlotvsdistance}}
\end{figure}
\begin{figure}[htbp]
\centering
\centering
\includegraphics[width=9.2cm]{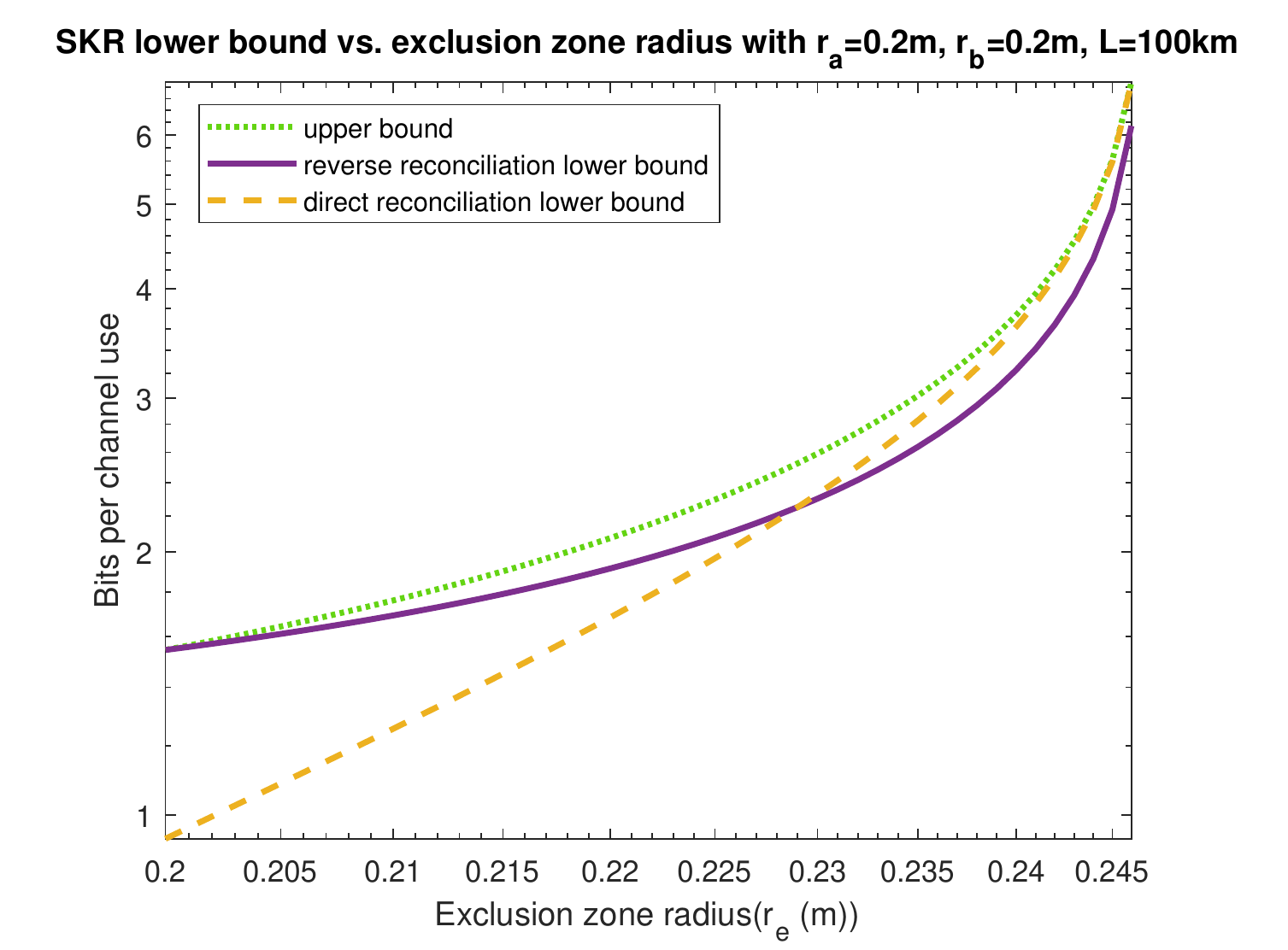}
\caption{SKR lower bound vs. exclusion zone radius. 
Wavelength $\lambda$ is set to 1550nm. Reconciliation efficiency is perfect ($\beta=1$) and the input power is set to infinity.\label{SKRvsExZo_cropped}}
\end{figure}

In Fig.~\ref{ExZo_Index202005012254_cropped} we can see that the optimal input power is infinity when $\beta=1$. 
Here in  Fig.~\ref{ExclusionZonePlotvsdistance} we plot the perfect reconciliation lower bounds as functions of transmission distance assuming equal aperture sizes for Alice and Bob with input power optimized (set to infinity). We also include upper bounds from~\cite{pan2019secret} with dotted curves. 
We can see from Fig.~\ref{ExclusionZonePlotvsdistance} that increasing the exclusion zone area can increase both direct and reverse reconciliation lower bounds, especially when the transmission distance $L$ is not too large. 
In Fig.~\ref{SKRvsExZo_cropped} we plot the upper bounds and lower bounds for both reconciliation schemes as functions of exclusion zone radius. It is clear that the SKR increment isn't linear to the increment of exclusion zone radius: when exclusion zone is larger, increasing the exclusion zone can help increase the SKR more. Also it's worth noticing that although direct reconciliation lower bound is lower than reverse reconciliation when no exclusion zone is set, it increases faster as the exclusion zone increases and eventually meets the upper bound. 
\begin{figure}[htbp]
\centering
\centering
\includegraphics[width=8.8cm]{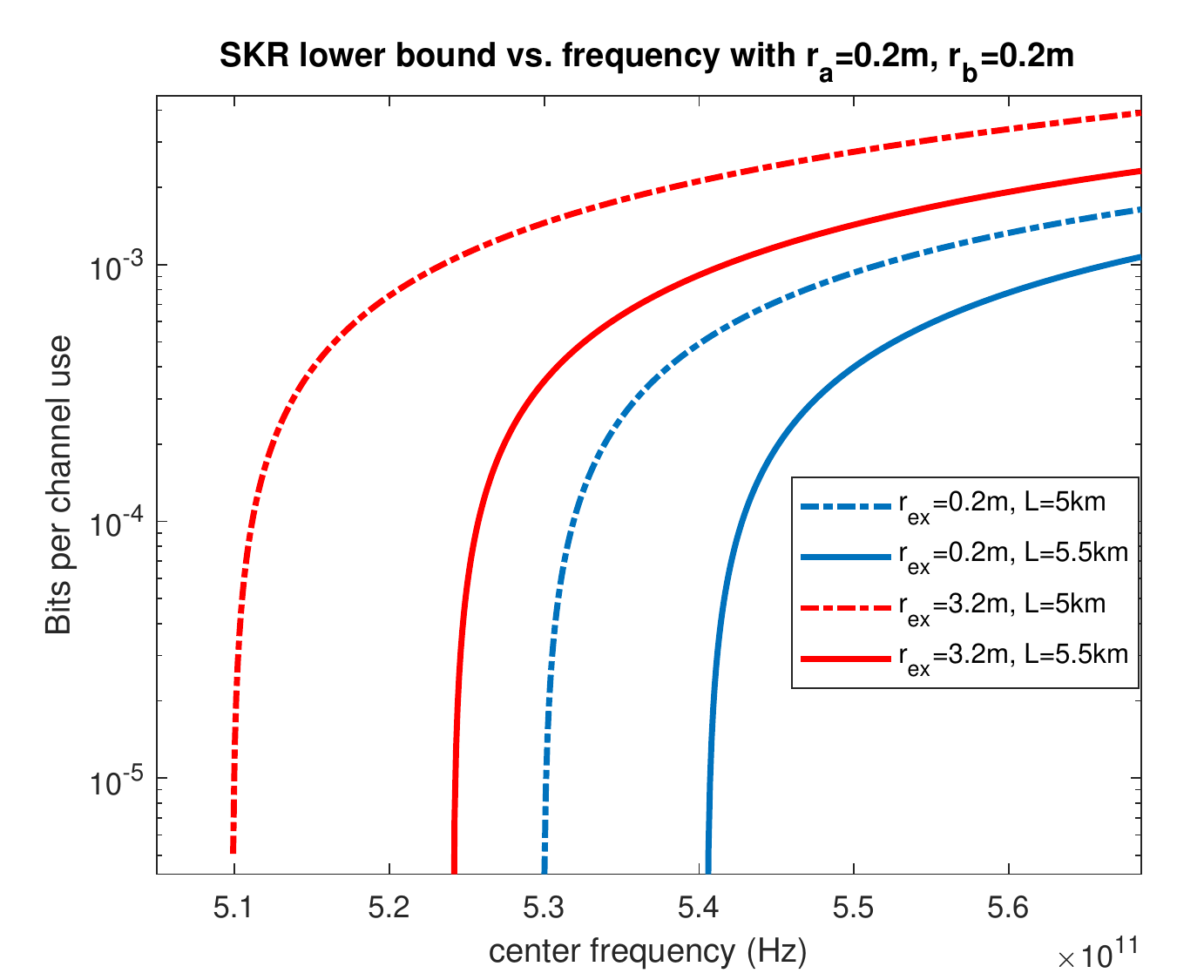}
\caption{SKR lower bound vs. center frequency. 
Radius of exclusion zone and transmission distance 
is used as a parameter. Reconciliation efficiency is perfect ($\beta=1$) and the input power is set to infinity. \label{ExclusionZonePlotvsfrequency}}
\end{figure}

In Fig.~\ref{ExclusionZonePlotvsfrequency} we plot the SKR lower bound versus the transmission center frequency, wherein the transmission distance $L$ and exclusion zone radius $r_{ex}$ are both used as parameters. Here solid curves and dash-dotted curves are with different transmission distances. 
It's worth noticing that the lower bounds in Fig.~\ref{ExclusionZonePlotvsfrequency} are attained by reverse reconciliation as direct reconciliation lower bound is zero under these parameters.

As we can see in Fig.~\ref{ExclusionZonePlotvsfrequency}, the SKR lower bounds increase with increasing frequency. 
Although choosing a higher frequency can always result in higher SKR, 
this can pose potential challenges to the system design as we need higher frequency for longer transmission distance. 
This downside can be mitigated 
by enlarging the exclusion zone as it effectively decreases Eve's receiving ability, relaxing the need for higher frequency as is illustrated in Fig.~\ref{ExclusionZonePlotvsfrequency} with red curves.

Next we analyze specific QKD protocols, Gaussian-modulated 
CVQKD protocol (with coherent states, heterodyne detection and reverse reconciliation) and Decoy State BB84 (DS-BB84) protocol and compare their performances under the exclusion zone scenario. We assume that Alice uses a weak coherent-state source and transmits signal-state pulses with on average $\mu$ photons per pulse at $R$ states/s. For CVQKD SKR we use the CCQ (classical-classical-quantum) rate (solid curve) as in Eq.~(69) from~\cite{pan2019secret} and for DS-BB84 (dotted curves) we use Eq.~(95) from~\cite{pan2019secret} with reconciliation efficiency $f_L$. Here we also include lower bounds obtained with Eqs.~(\ref{LBRmu}) as CQQ (classical-quantum-quantum) rate (dashed curves) as it doesn't assume a specific detection scheme for Bob.

\begin{figure}[htbp]
\centering
\centering
\includegraphics[width=8.8cm]{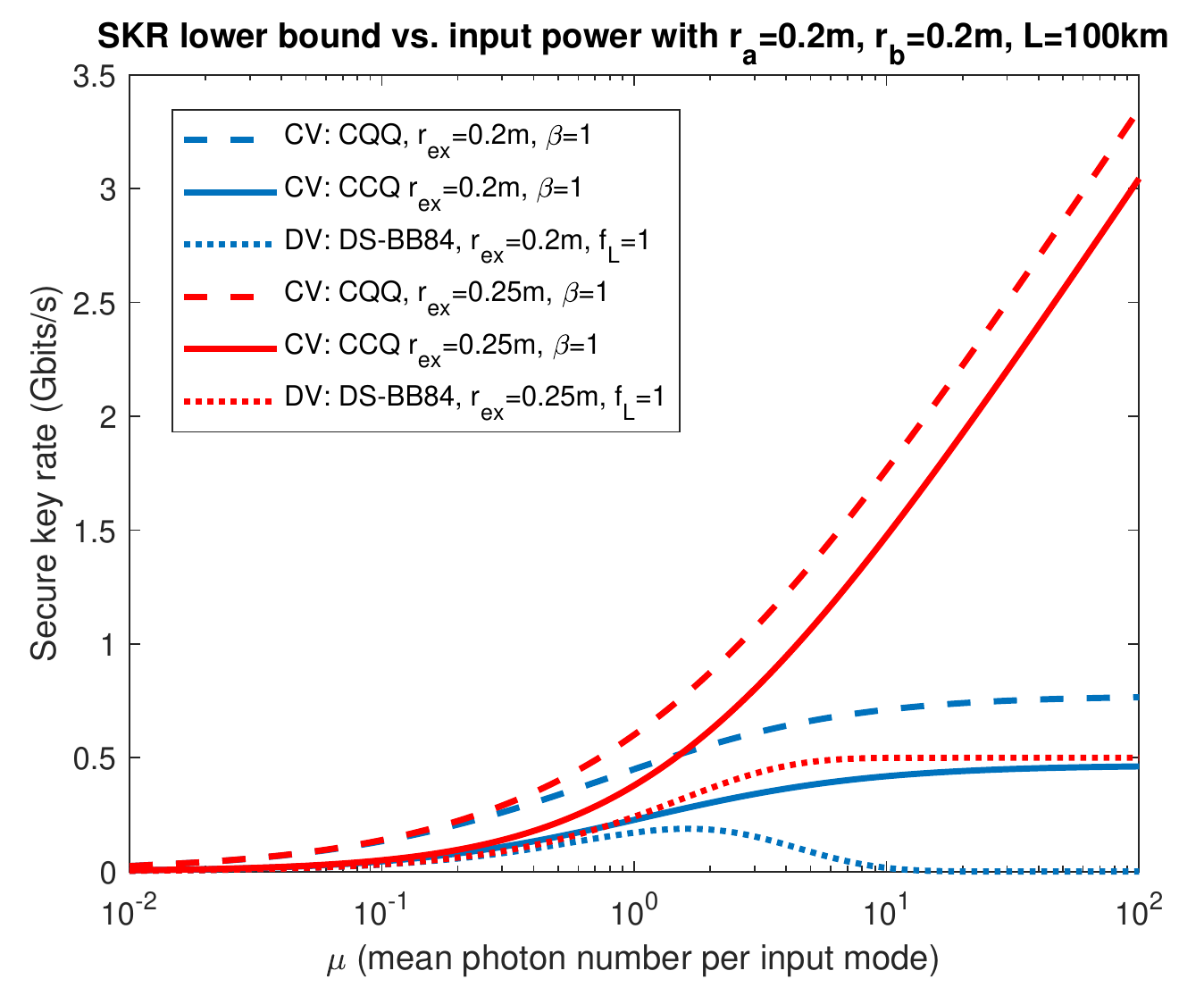}
\caption{SKR lower bound vs. input power with perfect information reconciliation. 
The transmission distance is 100km. Radius of exclusion zone ($r_{ex}$) is used as a parameter.  
Wavelength $\lambda$ is set to 1550nm. $R=$1Gbit/s. \label{CVvsDV_mu_cropped}}
\end{figure}

In Fig.~\ref{CVvsDV_mu_cropped} we plotted CQQ and CCQ rates for CVQKD and achievable rate for DS-BB84 versus input power. We can see that since we assume heterodyne detection for Bob instead of unspecified detection scheme the CCQ rate is upper bounded by CQQ rate. Both rates are largely increased under the exclusion zone ($r_{ex}=0.25m$) scenario. 

\begin{figure}[htbp]
\centering
\centering
\includegraphics[width=8.8cm]{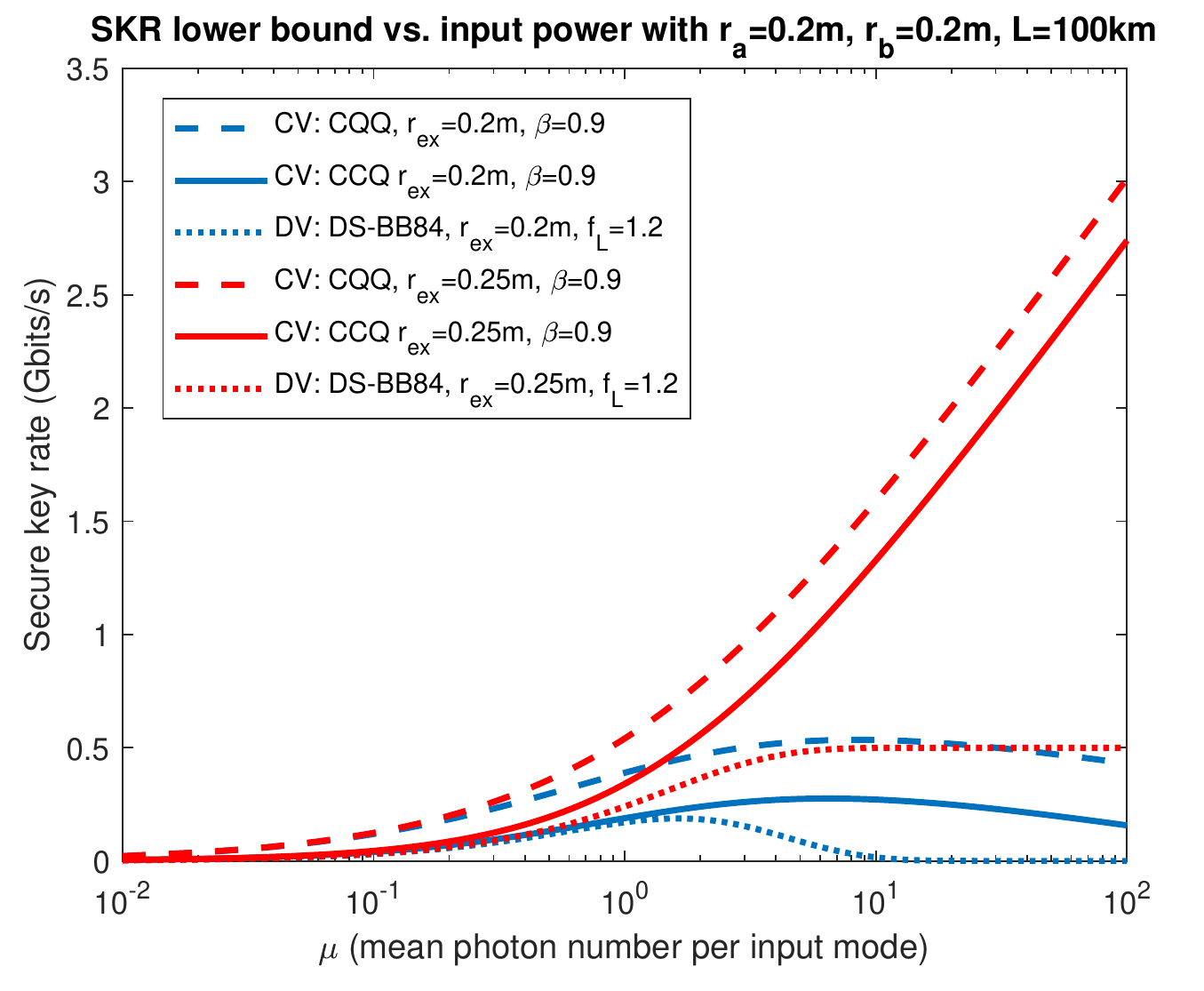}
\caption{SKR lower bound vs. input power with imperfect information reconciliation. 
The transmission distance is 100km. Radius of exclusion zone ($r_{ex}$) is used as a parameter.  
Wavelength $\lambda$ is set to 1550nm. $R=$1Gbit/s. \label{CVvsDVbeta9_mu_cropped}}
\end{figure}

Also here in Fig.~\ref{CVvsDV_mu_cropped} we can see that  DS-BB84 is upper bounded by CCQ rate. Although it has an optimal input power when there is no exclusion zone, setting one can increase its rate exceeding this optimal input power, achieving higher rates. In Fig.~\ref{CVvsDVbeta9_mu_cropped} we include the same comparison case with imperfect information reconciliation ($\beta=0.9, f_L=1.2$). We can see that imperfect information reconciliation renders finite optimal input power for both CCQ rate and DS-BB84 rate with similar comparison results while an exclusion zone ($r_{ex}$) still largely increases both CCQ rate and CQQ rate exceeding this finite optimal input power.

\begin{figure}[htbp]
\centering
\centering
\includegraphics[width=8.8cm]{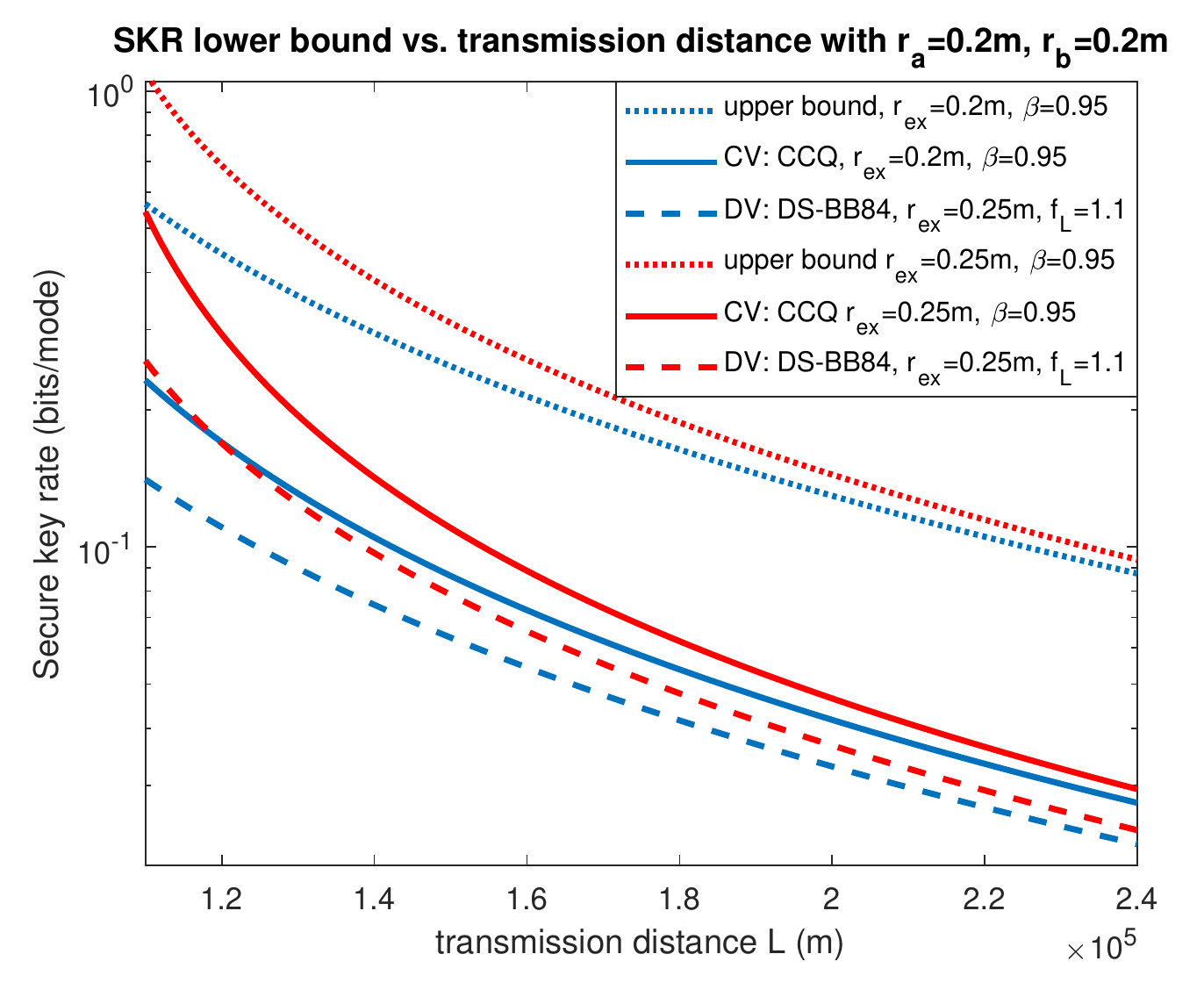}
\caption{SKR lower bounds vs. input power with imperfect information reconciliation. 
The transmission distance is 100km. Radius of exclusion zone ($r_{ex}$) is used as a parameter.  
Wavelength $\lambda$ is set to 1550nm. \label{CompVdistance_cropped}}
\end{figure}

\begin{figure}[htbp]
\centering
\centering
\includegraphics[width=8.8cm]{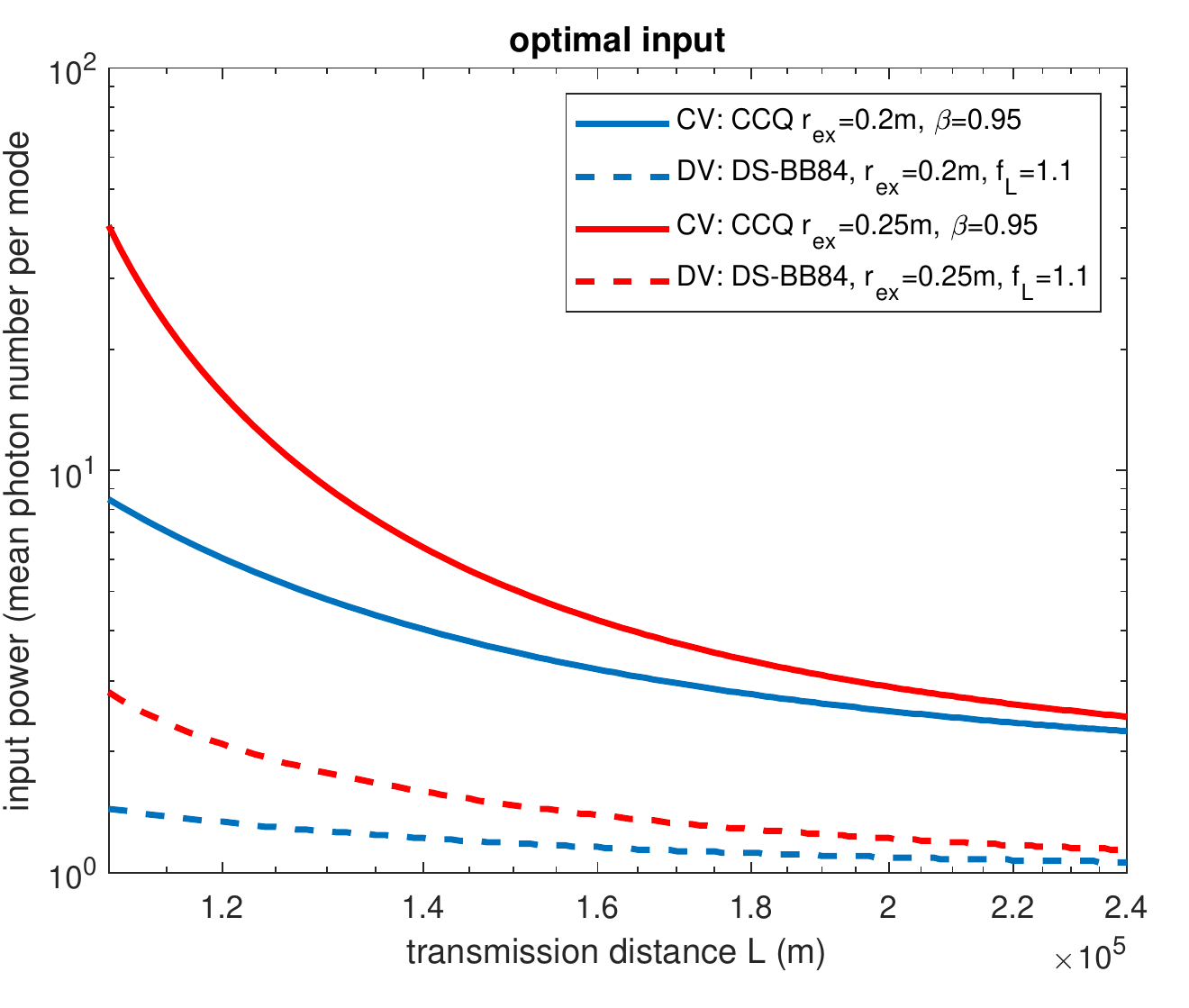}
\caption{Corresponding optimal input power for CCQ rate and DS-BB84 with or without exclusion zone in Fig.~\ref{CompVdistance_cropped}. 
\label{CompVdistanceOptimalInput_cropped}}
\end{figure}

 Below we plot the comparison for these three rates with optimized input power. Fig.~\ref{CompVdistance_cropped} captured the comparison against transmission distance for different exclusion zone sizes with corresponding optimal input power provided in Fig.~\ref{CompVdistanceOptimalInput_cropped}. We can see that the optimal input power decreases with increasing transmission distance.

\section{Limited aperture size combined with exclusion zone.}\label{CombinedScenario}

\begin{figure}[htbp] 
\centering{\includegraphics[height=11pc]{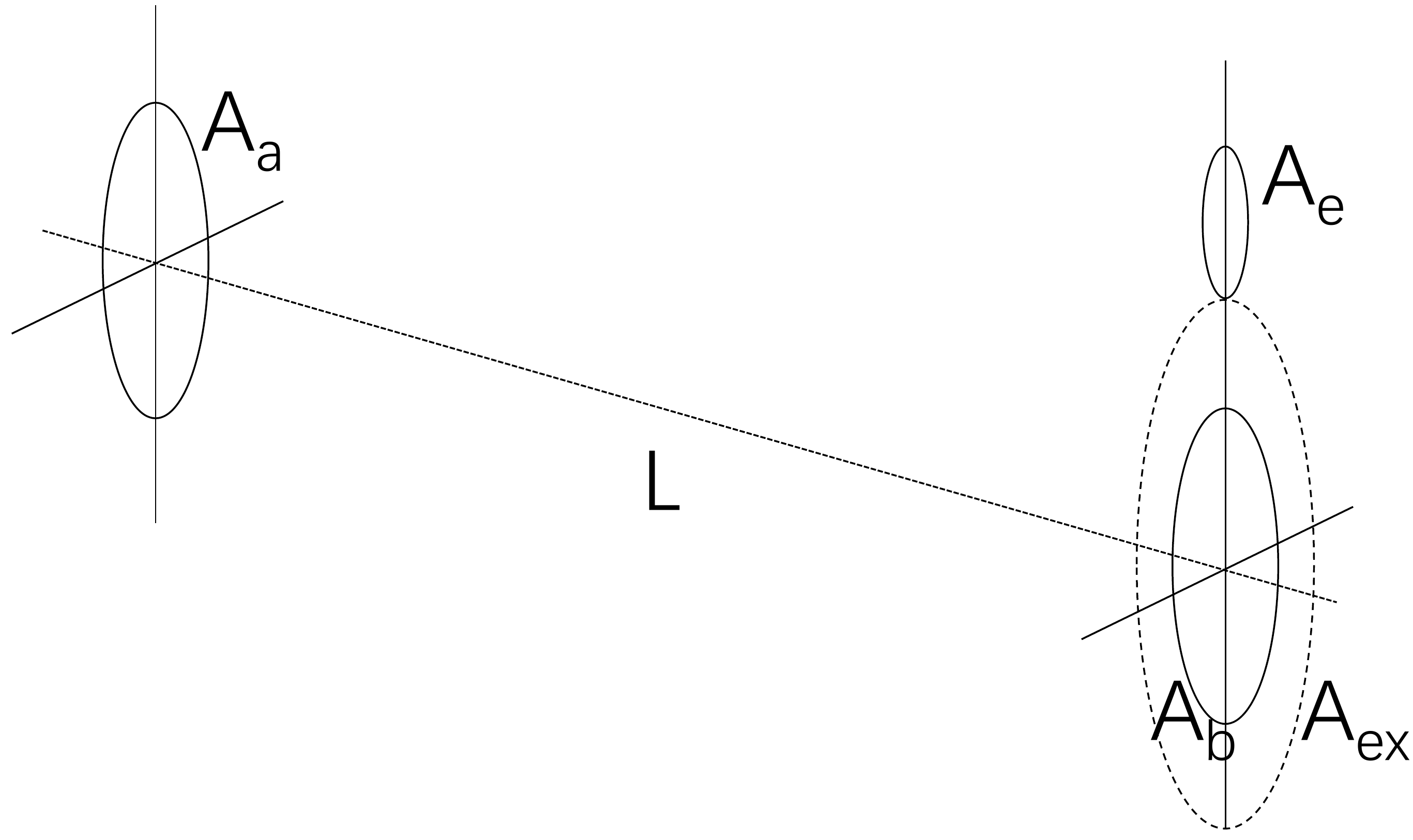}}
\caption{Geometric setup of limited-size aperture scenario. $A_a$ is the transmitting aperture (Alice) area and $A_b$ is the receiving aperture (Bob) area. $L$ is the  transmission distance between Alice and Bob. $A_{e}$ is the eavesdropper (Eve) aperture area which is placed in the same plane as Bob's aperture. The exclusion zone area  is denoted as $A_{ex}$ which is a circular area centered at Bob's aperture. \label{ExclusionZoneLimitedAperture}}
\end{figure}

\noindent In this section we combine the exclusion zone scenario with the limited-sized aperture scenario where Eve's aperture size is limited~\cite{pan2020ICTON} and investigate how it can improve security as a defense strategy. First we introduce the problem setup as is illustrated in Fig.~\ref{ExclusionZoneLimitedAperture}. Similar to Fig.~\ref{FSO}, here $A_a$, $A_b$, $A_{ex}$ respectively denote the area of transmitter aperture (Alice), receiver aperture (Bob) and the exclusion zone. Additionally we have a limited-sized eavesdropper aperture in the same plane of Bob, denoted as $A_e$. In this case, Eve's optimal position is shown in Fig.~\ref{ExclusionZoneLimitedAperture} to be tangential to the exclusion zone.

We assume Gaussian beam (transverse coordinates $x$, $y$) with beam waist radius $W_0=r_a$ transmitted with propagation distance $L$ and intensity at the center of the beam waist $I_0$: 
\begin{align}
\|U(\rho,L)\|^2&=I_0\left(\frac{W_0}{W(L)}\right)^2\exp\left({-\frac{2\rho^2}{W^2(L)}}\right),\\
    z_0&=W_0^2\frac{\pi}{\lambda},\\
    W(L)&=W_0\sqrt{1+(L/z_0)^2},\\
    \rho^2&=x^2+y^2.
\end{align}
The total power in this Gaussian beam is:
\begin{equation}
    P_{total}=I_0\frac{\pi  W_0^2}{2}\label{Ptotal},
\end{equation}

Without loss of generality we can assume Eve's aperture right above the exclusion zone as in Fig.~\ref{ExclusionZoneLimitedAperture} and derive Bob's received power ($P_{Bob}$) and Eve's received power ($P_{Eve}$) below.

\begin{align}
    P_{Bob}&=I_0\int_{-r_b}^{r_b} \frac{\pi ^{3/2} W_0^3 e^{-\frac{2 \pi ^2 W_0^2 y^2}{A}} \text{erf}\left(\frac{\sqrt{2} \pi  W_0 \sqrt{r_b^2-y^2}}{\sqrt{A}}\right)}{\sqrt{2} \sqrt{A}} \, dy\label{PBob}\\
    P_{Eve}&=\nonumber\\
    &I_0\int_{-r_e}^{r_e} \frac{-\pi ^{3/2} W_0^3 e^{-\frac{2 \pi ^2 W_0^2 x^2}{A}} E\left(\frac{\sqrt{2} \pi  W_0 C}{\sqrt{A}},\frac{\sqrt{2} \pi  W_0 B}{\sqrt{A}}\right)}{2 \sqrt{2} \sqrt{A}} \, dx\label{PEve}
\end{align}

with 

\begin{align}
    A&=\pi ^2 W_0^4+\lambda ^2 L^2\\
    B&=\sqrt{r_e^2-x^2}\\
    C&=r_{ex}+r_e\\
    E(x,y)&=\text{erf}\left(x-y\right)-\text{erf}\left(x+y\right)
\end{align}

With Eqs.~(\ref{Ptotal}) - (\ref{PEve}), Alice-to-Bob transmissivity ($\eta$) and Eve's fraction of collected power ($\kappa$) can be expressed as below: 
\begin{align}
    \eta&=\frac{P_{Bob}}{P_{total}},\label{eta}\\
    \kappa&=\frac{P_{Eve}}{(1-\eta)P_{total}}\label{kappa},
\end{align}
Also for noise frequency dependence we use the black body radiation equation Eq.~(\ref{blackradia}).

With the above expressions, first we plot the input power dependency for different transmission distances. Below in Fig.~\ref{InputDependencybeta1_10km_cropped} we plot the direct and reverse reconciliation achievable rate lower bounds as functions of input power for different exclusion zone radii. Here the radii of Alice's, Bob's and Eve's aperture are all set to 5cm. The transmission is over 10km with reconciliation efficiency $\beta=1$. Similar to Fig.~\ref{ExZo_Index202005012254_cropped} we can see that the optimal input power here is infinity. Not surprisingly, by setting an exclusion zone ($r_{ex}=0.5$m) we can increase the achievable rates for both reconciliation schemes. We can also see that direct reconciliation rate is higher than reverse reconciliation when transmission distance is small ($\eta$ is large). Since most power is focused at the center of the transmitted Gaussian beam, which is mostly received by Bob at line of sight, the corresponding $\kappa$ is small. This explains why direct reconciliation scheme achieves higher rate than reverse reconciliation scheme. 

\begin{figure}[htbp]
\centering
\centering
\includegraphics[width=8.8cm]{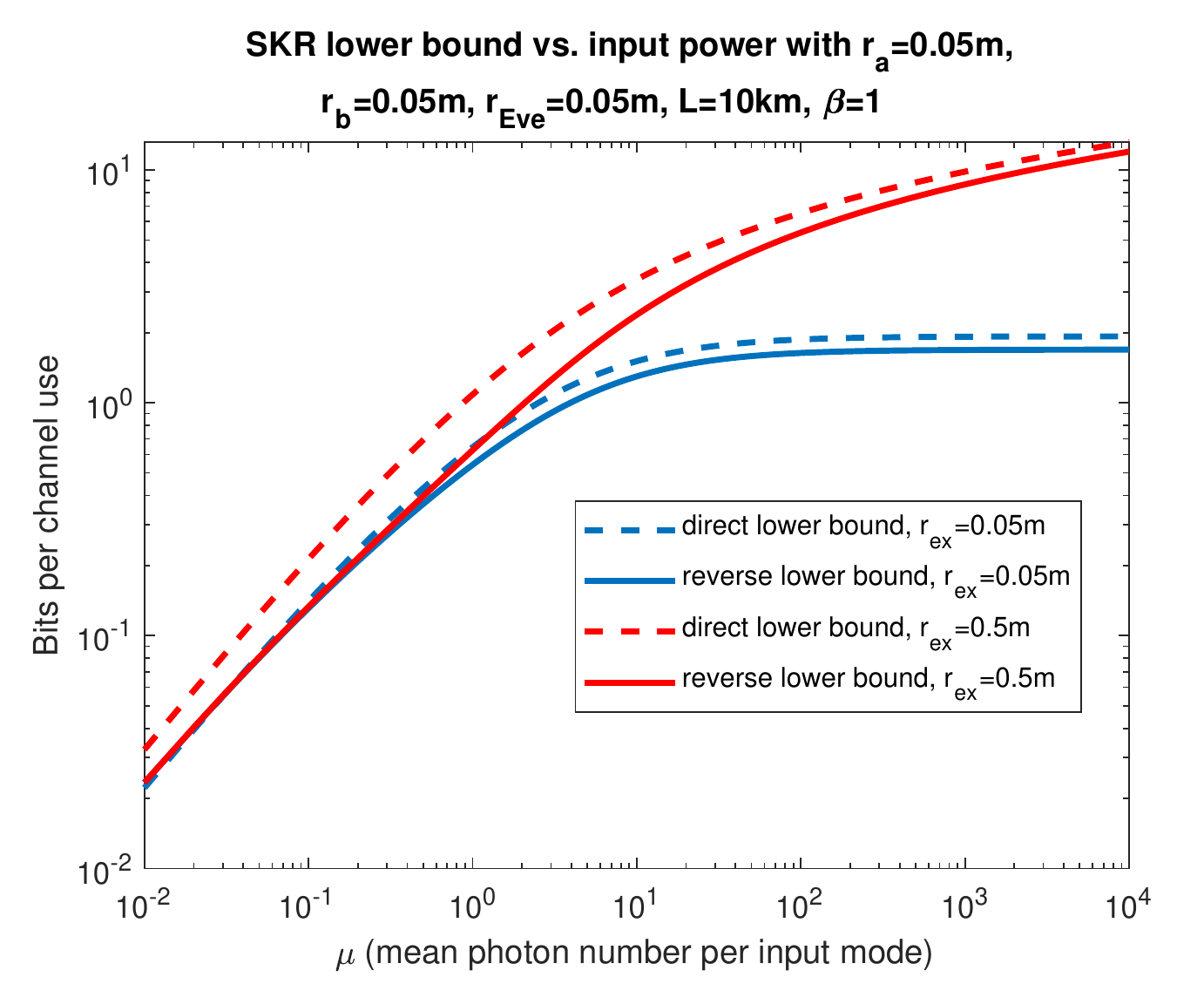}
\caption{SKR lower bounds vs. input power. 
The transmission distance $L$ is 10km. Radius of exclusion zone  ($r_{ex}$) is specified in the legend. Reconciliation efficiency $\beta$ is set to 1. Transmission center wavelength $\lambda$ is set to 1550nm. Transmitted Gaussian beam waist radius is set to $W_0=5$cm. Bob, Alice and Eve aperture radius are set to $r_b=r_a=r_{Eve}=5$cm. \label{InputDependencybeta1_10km_cropped}}
\end{figure}

\begin{figure}[htbp]
\centering
\centering
\includegraphics[width=8.8cm]{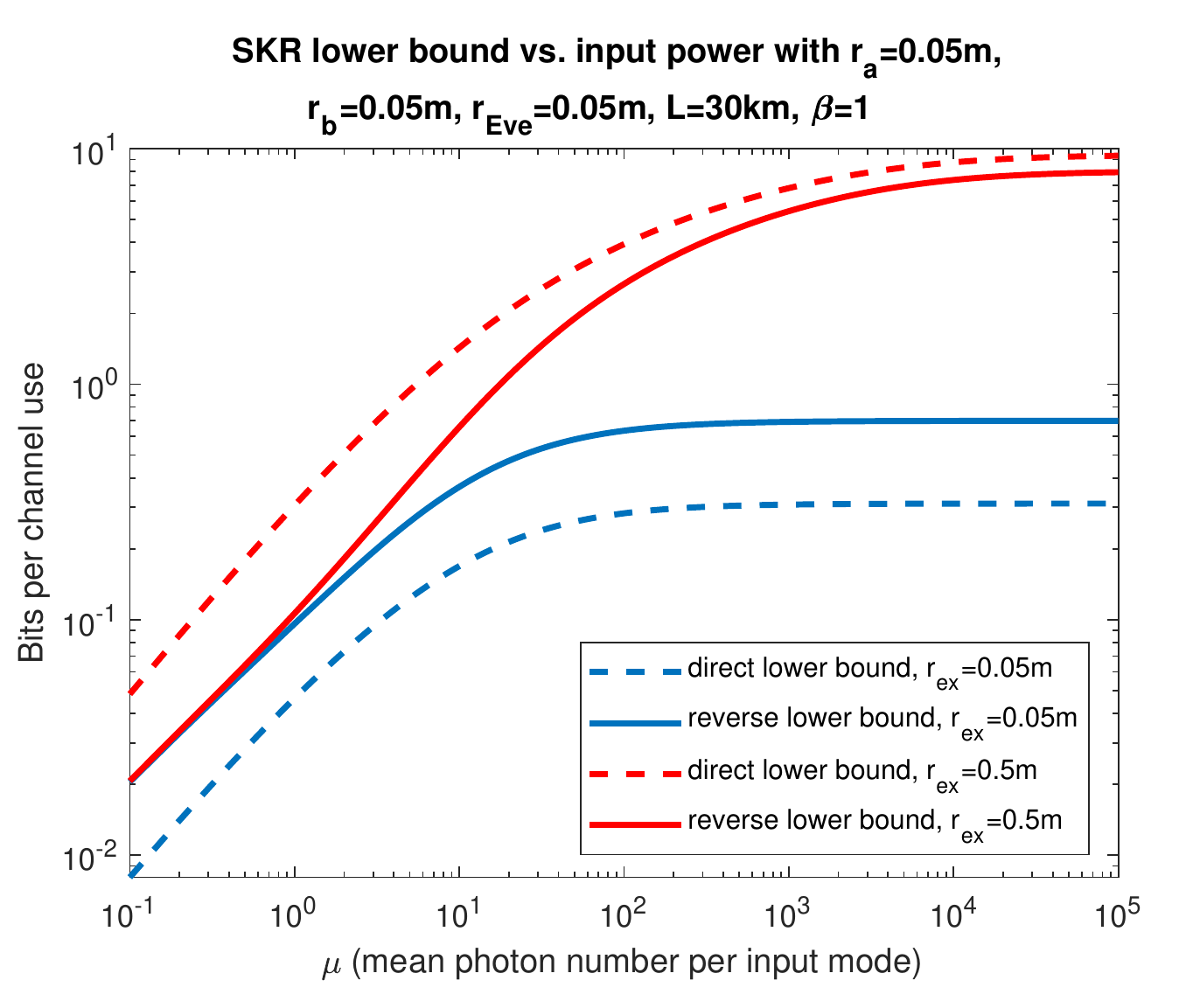}
\caption{SKR lower bounds vs. input power. 
The transmission distance $L$ is 30km. Radius of exclusion zone  ($r_{ex}$) is specified in the legend. Reconciliation efficiency $\beta$ is set to 1. Transmission center wavelength $\lambda$ is set to 1550nm. Transmitted Gaussian beam waist radius is set to $W_0=5$cm. Bob, Alice and Eve aperture radius are set to $r_b=r_a=r_{Eve}=5$cm. \label{InputDependencybeta1_30km_cropped}}
\end{figure}

\begin{figure}[htbp]
\centering
\centering
\includegraphics[width=8.8cm]{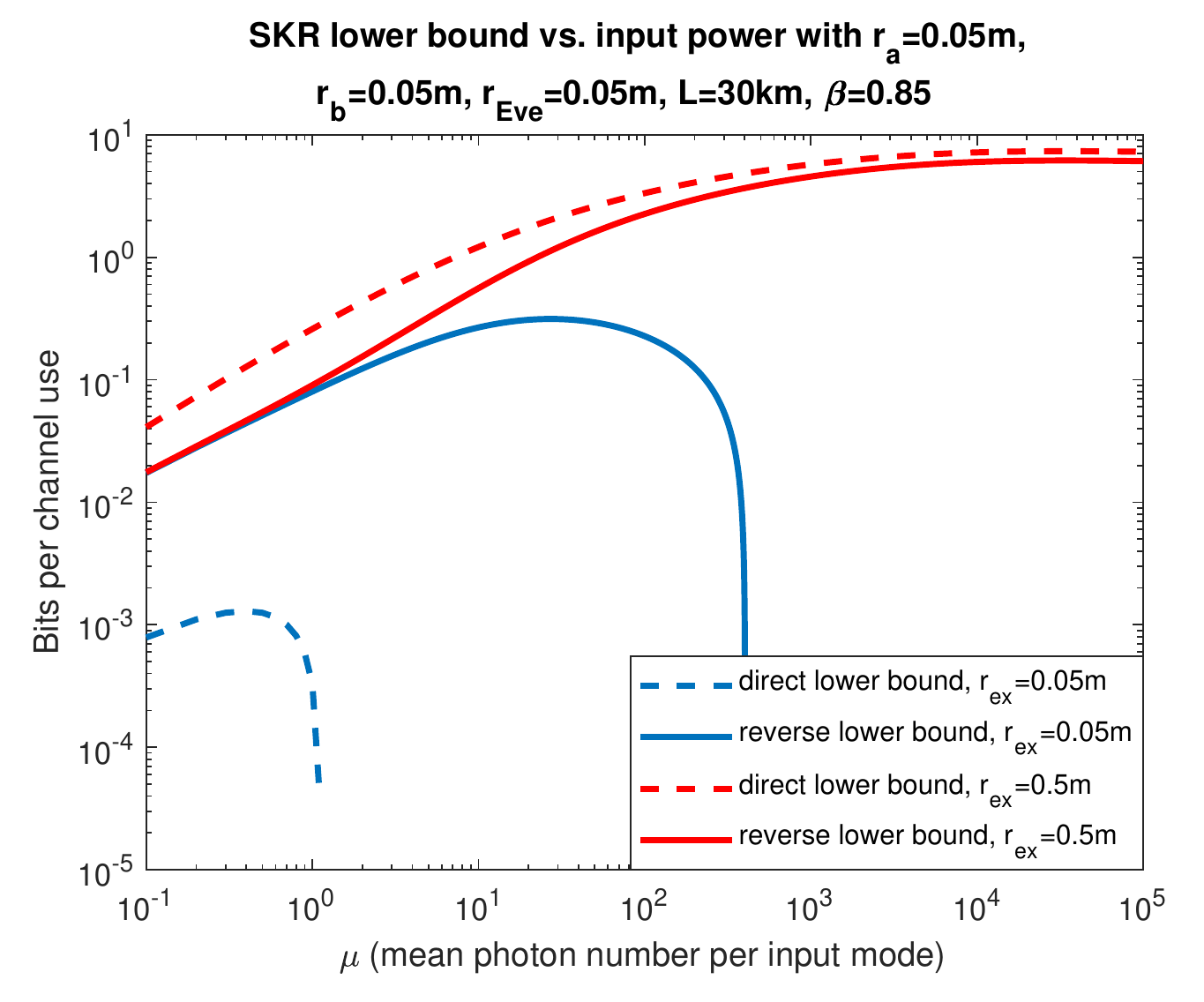}
\caption{SKR lower bounds vs. input power. 
The transmission distance $L$ is 30km. Radius of exclusion zone  ($r_{ex}$) is specified in the legend. Reconciliation efficiency $\beta$ is set to 1. Transmission center wavelength $\lambda$ is set to 1550nm. Transmitted Gaussian beam waist radius is set to $W_0=5$cm. Bob, Alice and Eve aperture radius are set to $r_b=r_a=r_{Eve}=5$cm. \label{InputDependencybeta_85_30km_cropped}}
\end{figure}

However, when we increase the transmission distance, as is presented in Fig.~\ref{InputDependencybeta1_30km_cropped} here with only the transmission distance increased to 30km, we can see that with no exclusion zone set ($r_{ex}=5$cm), direct reconciliation is more sensitive to the consequential channel loss increment and performs worse than the reverse reconciliation lower bound. Yet if an exclusion zone is set ($r_{ex}=0.5$m), compared to the reverse reconciliation, the direct reconciliation benefits more from this strategy specifically designed to decrease $\kappa$ and we can see that in this case the direct reconciliation lower bound increases to be higher than the reverse reconciliation.

This is even more evident if we set the reconciliation efficiency to be imperfect, which is mostly the case in realistic implementation. In Fig.~\ref{InputDependencybeta_85_30km_cropped} we set reconciliation efficiency $\beta$ to 0.85. As a result when no exclusion zone is set ($r_{ex}=0.5$m), the optimal input power is shown to decrease from infinity to some finite values. Aside from that, we can see that an imperfect reconciliation process decreases the direct reconciliation rate much more than it does to the reverse reconciliation. However, when an exclusion zone is set ($r_{ex}=0.5$m), still the direct reconciliation rate is  higher than the reverse reconciliation rate, exceeding the SKR results corresponding to the original optimal input power.

\begin{figure}[htbp]
\centering
\includegraphics[width=8.8cm]{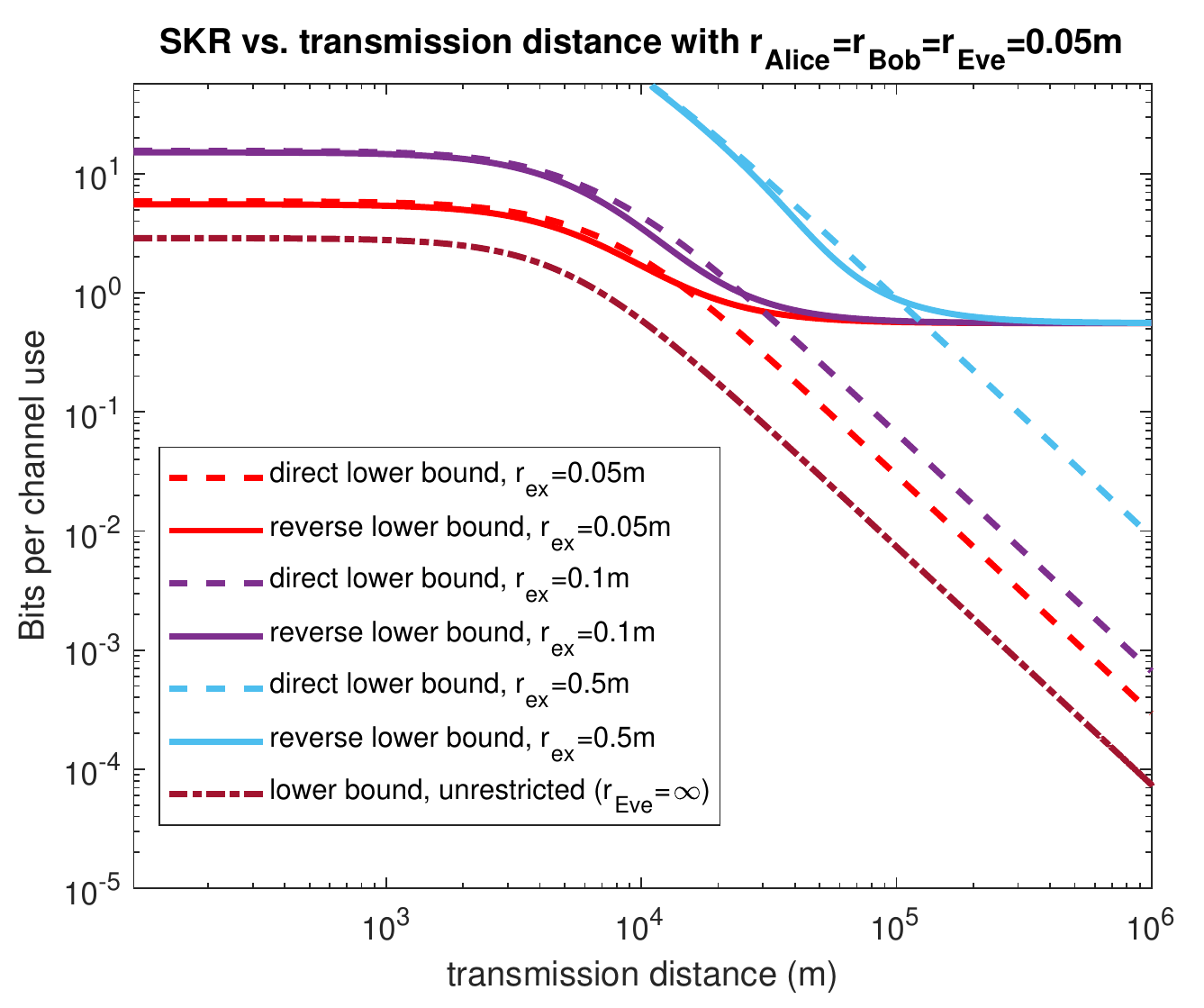}
\caption{SKR lower bounds vs. distance. 
Unrestricted Eve's case~\cite{pan2019secret} is also included for comparison. \label{Index202008042219}}
\end{figure}

Next we look into the perfect reconciliation case with optimized input power. In Fig.~\ref{Index202008042219} we plotted the direct and reverse reconciliation lower bounds as functions of transmission distance for different exclusion zone radii. Here the unrestricted Eve's case is also included for comparison. We can see that lower bounds exceeding the unrestricted Eve's case can be achieved. Here the direct reconciliation lower bounds exceed the reverse reconciliation lower bounds when the transmission distance is not too large. When the transmission distance increases, the direct reconciliation lower bounds start to decrease in a similar manner to the unrestricted Eve's case while the reverse reconciliation lower bounds converge to a constant rate as we saw in our previous work~\cite{pan2020ICTON}. We can see that an exclusion zone mostly helps increase the achievable rate when the transmission distance is not too large. For example, when the exclusion zone radius is large ($r_{ex}=0.5$m) the achievable rate even goes to infinity when transmission distance is over 10km, exceeding the constant rate convergence at short transmission distance. We can also see that with a larger exclusion zone, the achievable rate at a larger transmission distance can be increased. Yet this doesn't change the constant rate convergence as transmission distance goes to infinity since an exclusion zone, when its radius is comparable to the communicating parities' aperture radii,  doesn't change the ratio of Eve's collecting ability versus Bob's collecting ability too much at large transmission distance.

\begin{figure}[htbp]
\centering
\includegraphics[width=8.8cm]{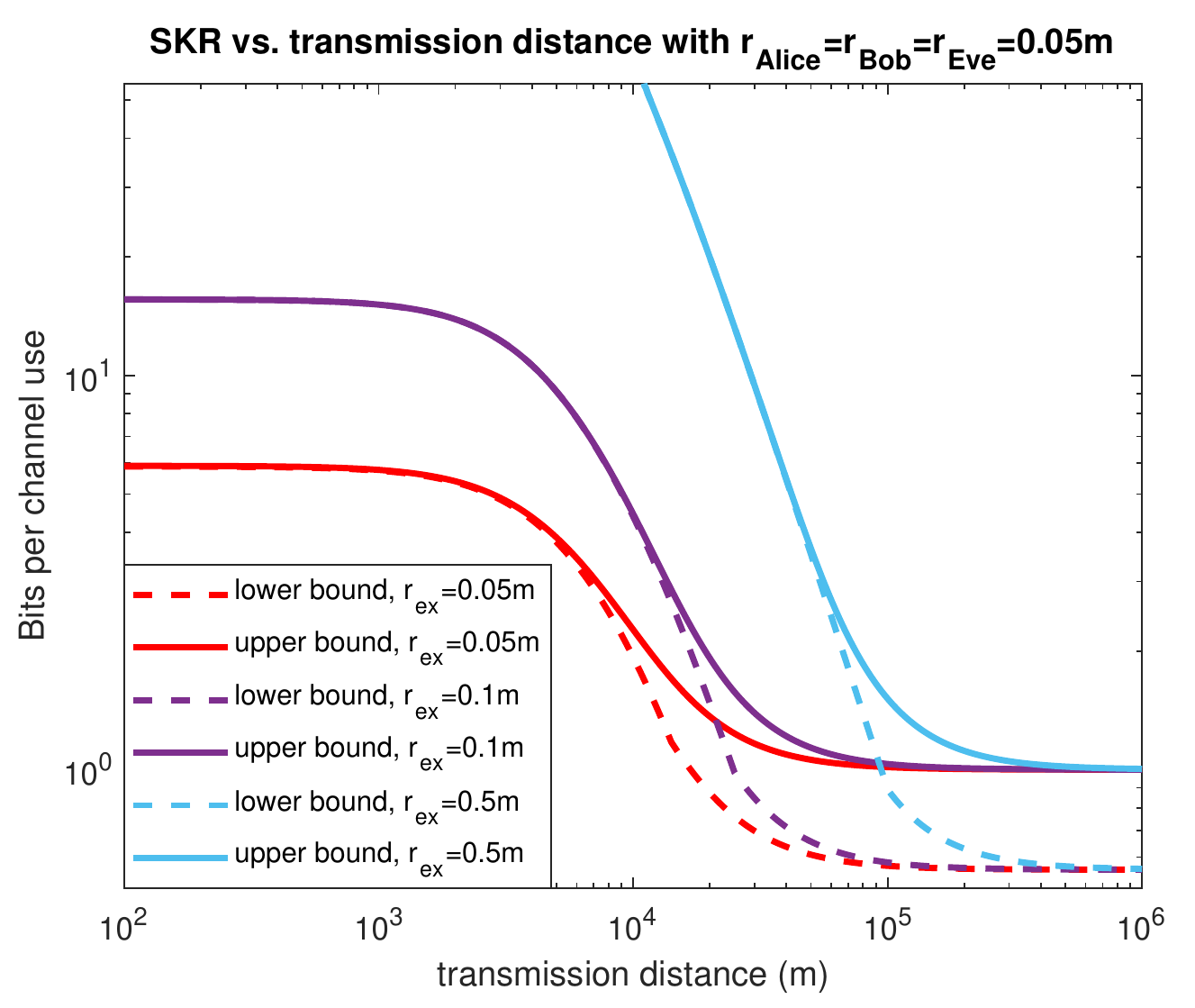}
\caption{SKR lower bounds and upper bounds vs. distance. 
 \label{Index202008111617}}
\end{figure}

In Fig.~\ref{Index202008111617} we plot the lower bounds as the maximum of direct and reverse lower bounds versus the upper bounds as functions of the transmission distance. Here we still use the same parameters as in Fig.~\ref{Index202008042219}. We can see that when transmission distance is not too large, the upper bounds closely match the lower bounds, giving the capacity in this region. 

\begin{figure}[htbp]
\centering
\includegraphics[width=8.8cm]{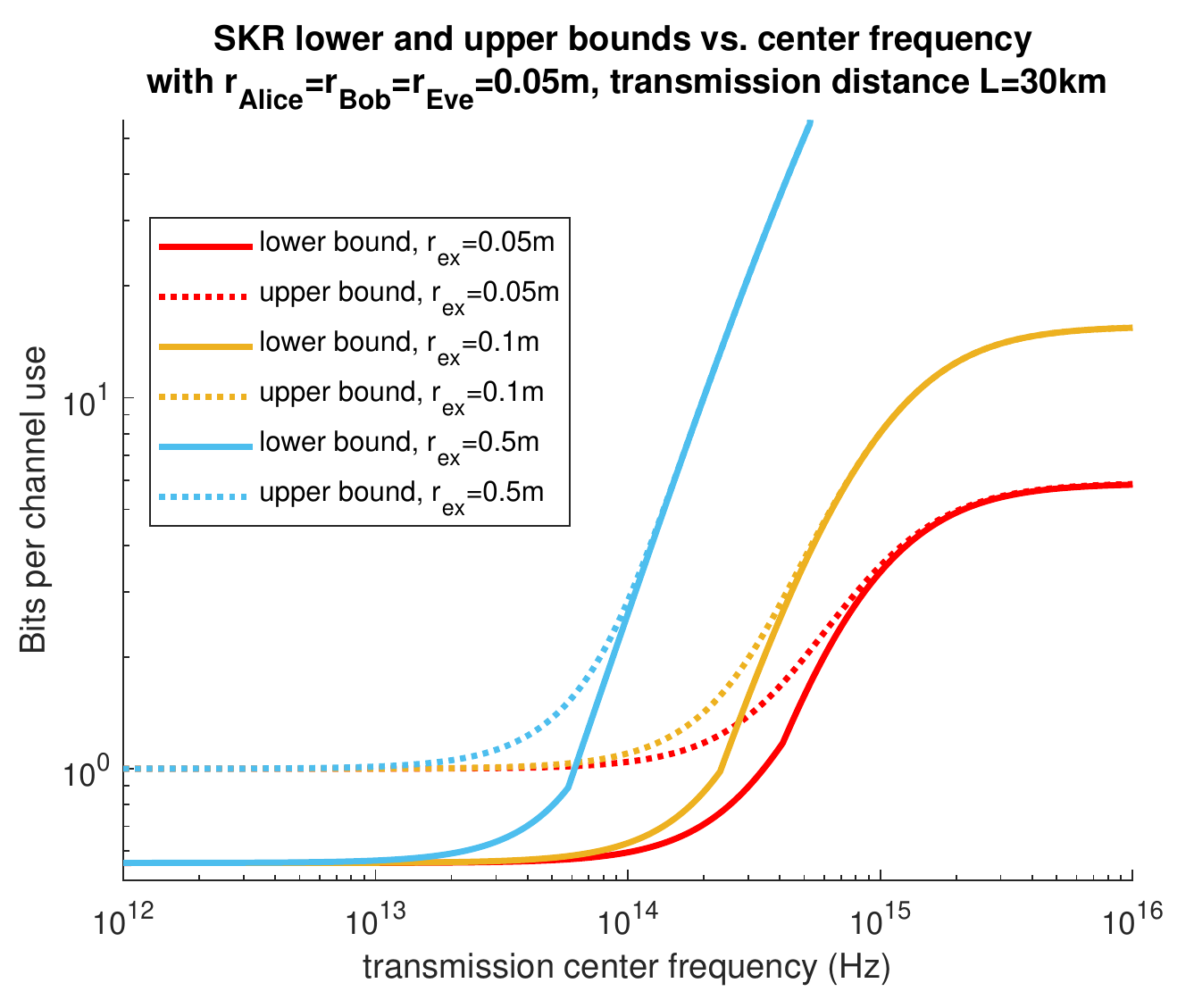}
\caption{SKR lower bounds and upper bounds vs. frequency. 
 \label{Index202008112147}}
\end{figure}

In Fig.~\ref{Index202008112147} we plot the lower bounds versus upper bounds as functions of the transmission center frequency. We can see that an exclusion zone can help increase the achievable rate especially when the transmission frequency is high. When the exclusion zone radius increases, similarly to Fig.~\ref{ExclusionZonePlotvsfrequency}, this helps to achieve a higher achievable rate with low transmission center frequency. It's worth noticing that when $r_{ex}=0.5$m, the achievable rate even goes to infinity instead of converging to a constant rate when the exclusion zone is large.

\section{Summary}


\noindent In this paper, we have analyzed SKR lower bounds and upper bounds for the realistic scenario in free-space satellite-to-satellite secure communication where an exclusion zone is set around the legitimate receiver and studied the performance with respect to relevant channel parameters. By enlarging the exclusion zone area we can relax the need for high frequency in long distance transmission and achieve a higher secure key rate. 
When Eve's aperture size is limited, an exclusion zone can serve as a good defense strategy to increase the secure key rate. 

\section*{Acknowledgement}
This paper was supported in part by L3Harris, MURI ONR, NSF, and GD. The authors thankfully acknowledge helpful discussions with Saikat Guha, Kaushik Seshadreesan and John Gariano from the University of Arizona, Jeffrey Shapiro from Massachusetts Institute of Technology and William Clark, Mark R. Adcock from General Dynamics.

\bibliographystyle{IEEEtran}
\bibliography{QSDC}






\end{document}